\documentclass[3p,times,procedia]{elsarticle}
\usepackage{nupha_ecrc}

\volume{00}

\firstpage{1}

\journalname{Nuclear Physics A}

\runauth{}

\jid{nupha}

\jnltitlelogo{Nuclear Physics A}

\usepackage{amssymb}

\usepackage[figuresright]{rotating}

\begin{document}
	
	\begin{frontmatter}
		
		
		
		\dochead{XXVIIIth International Conference on Ultrarelativistic Nucleus-Nucleus Collisions\\ (Quark Matter 2019)}
		\title{Light Nuclei ($d$, $t$) Production in Au + Au Collisions at \\ $\sqrt{s_{NN}}$ = 7.7 - 200 GeV}
		
		
		\author{Dingwei \textsc{Zhang}$^{1}$ (for the STAR Collaboration)}
		
		\address{$^{1}$Key Laboratory of Quark $\&$ Lepton Physics (MOE) and Institute of Particle Physics, Central China Normal University, Wuhan, 4300    79, China }
		
		\begin{abstract}
			In high-energy nuclear collisions, light nuclei can be regarded as a cluster of baryons and their yields are sensitive to the baryon density fluctuations. Thus, the production of light nuclei can be used to study the QCD phase transition, at which the baryon density fluctuation will be enhanced. A yield ratio of light nuclei, defined as $N(t)$$\times$$N(p)$/$N^2(d)$, is predicted to be a sensitive observable to search for the 1st-order phase transition and/or QCD critical point in heavy-ion collisions. In this paper, we present the energy and centrality dependence of (anti)deuteron and triton production in Au+Au collisions at $\sqrt{s_{\mathrm{NN}}}$  =  7.7, 11.5, 14.5, 19.6, 27, 39, 54.4, 62.4, and 200 GeV measured by the STAR experiment at RHIC. We show beam-energy dependence for the coalescence parameter, $B_2(d)$ and $B_3(t)$, particle ratios, $d/p$, $t/p$, and $t/d$, and the yield ratio of $N(t)$$\times$$N(p)$/$N^2(d)$. More importantly, non-monotonic energy dependence is observed for the yield ratio, $N(t)$$\times$$N(p)$/$N^2(d)$, in 0-10\% central Au+Au collisions with a peak around 20-30 GeV. Their physics implications on QCD critical point search and change of the equation of state will be discussed.
			
		\end{abstract}
		
		
	\end{frontmatter}
	
	
	\section{Introduction}
	In relativistic heavy-ion collisions, the study of the phase transition between the Quark-Gluon Plasma (QGP) and the hadronic matter attracts great interest~\cite{Fukushima:2013rx}. Production of light nuclei with small binding energy is sensitive to the local nucleon density, which can provide a unique tool to probe the essential feature of the QCD phase diagram~\cite{Gutbrod:1988gt}. In the coalescence picture, the density of the cluster is proportional to the proton density times the probability of finding a neutron within a small sphere around the proton momentum~\cite{Csernai:1986qf}. Nucleon coalescence mechanism can be described as: 
	\begin{equation}
	{E_A}\frac{{{d^3}{N_A}}}{{{d}{p^3_A}}} = {B_A}{\left( {{E_p}\frac{{{d^3}{N_p}}}{{{d}{p^3_p}}}} \right)^Z}{\left( {{E_n}\frac{{{d^3}{N_n}}}{{{d}{p^3_n}}}} \right)^{A - Z}} \approx {B_A}{\left( {{E_p}\frac{{{d^3}{N_p}}}{{{d}{p^3_p}}}} \right)^A},
	\end{equation}
	where $A$ and $Z$ are the mass and charge number of the light nucleus under study. $p_p$, $p_n$, and $p_A$ are momenta of proton, neutron, and nucleus respectively, with $p_A = Ap_p$, assuming $p_p \approx p_n$.
	
	In the vicinity of the critical point or the first order phase transition, density fluctuations become larger. Based on the coalescence picture, the yield ratio, $N(t)$$\times$$N(p)$/$N^2(d)$, is sensitive to the neutron density fluctuation, $\Delta n$=$\langle(\delta n)^2\rangle/\langle n\rangle^2$, where $\langle n\rangle$ denotes the average value over space and $\delta n$ denotes the fluctuation of neutron density from its average value $\langle n\rangle$~\cite{Sun:2018jhg}. The yield ratio of light nuclei, which connects neutron density fluctuation, can be approximated as: 
	\begin{equation}
	N(t) \times N(p)/N^2(d)=g(1+\Delta n), 
	\end{equation} 
	with\ g = 0.29. Experimentally, one can measure the light nuclei yield ratio to probe the QCD critical point or first order phase transition.
	
	\section{Results and discussions}
	\subsection{Transverse momentum spectra}
	The results presented in this paper are obtained from the data taken with the STAR experiment in Au+Au collisions at $\sqrt{s_\mathrm{NN}}=$7.7-200 GeV at RHIC. The main detectors used in this analysis are the Time Projection Chamber (TPC)~\cite{Anderson:2003ur} and Time-Of-Flight (TOF)~\cite{Llope:2005yw}. 
	\begin{figure}[htpb]
		\centering
		\includegraphics[width=0.496\textwidth]{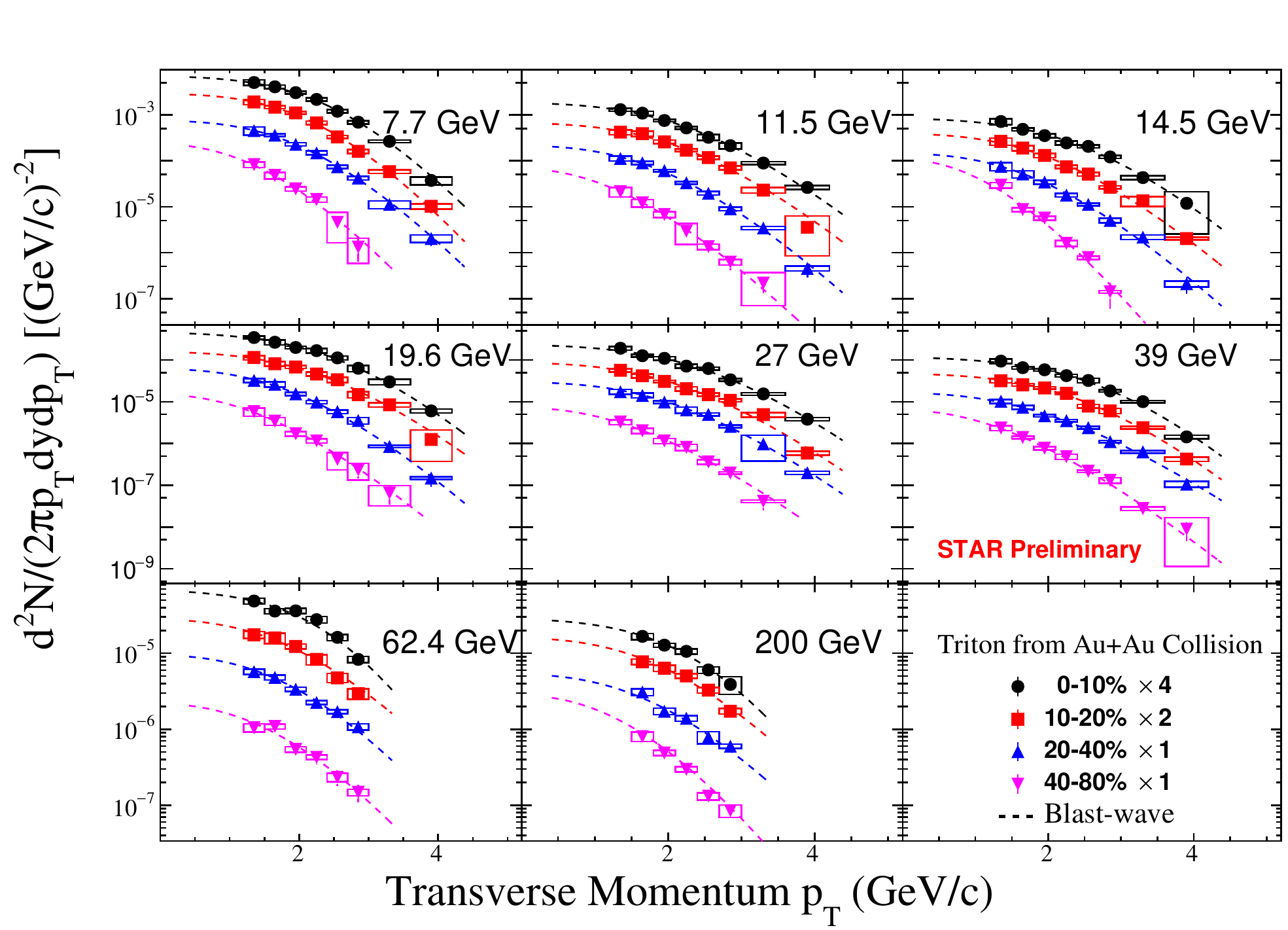}
		\includegraphics[width=0.496\textwidth]{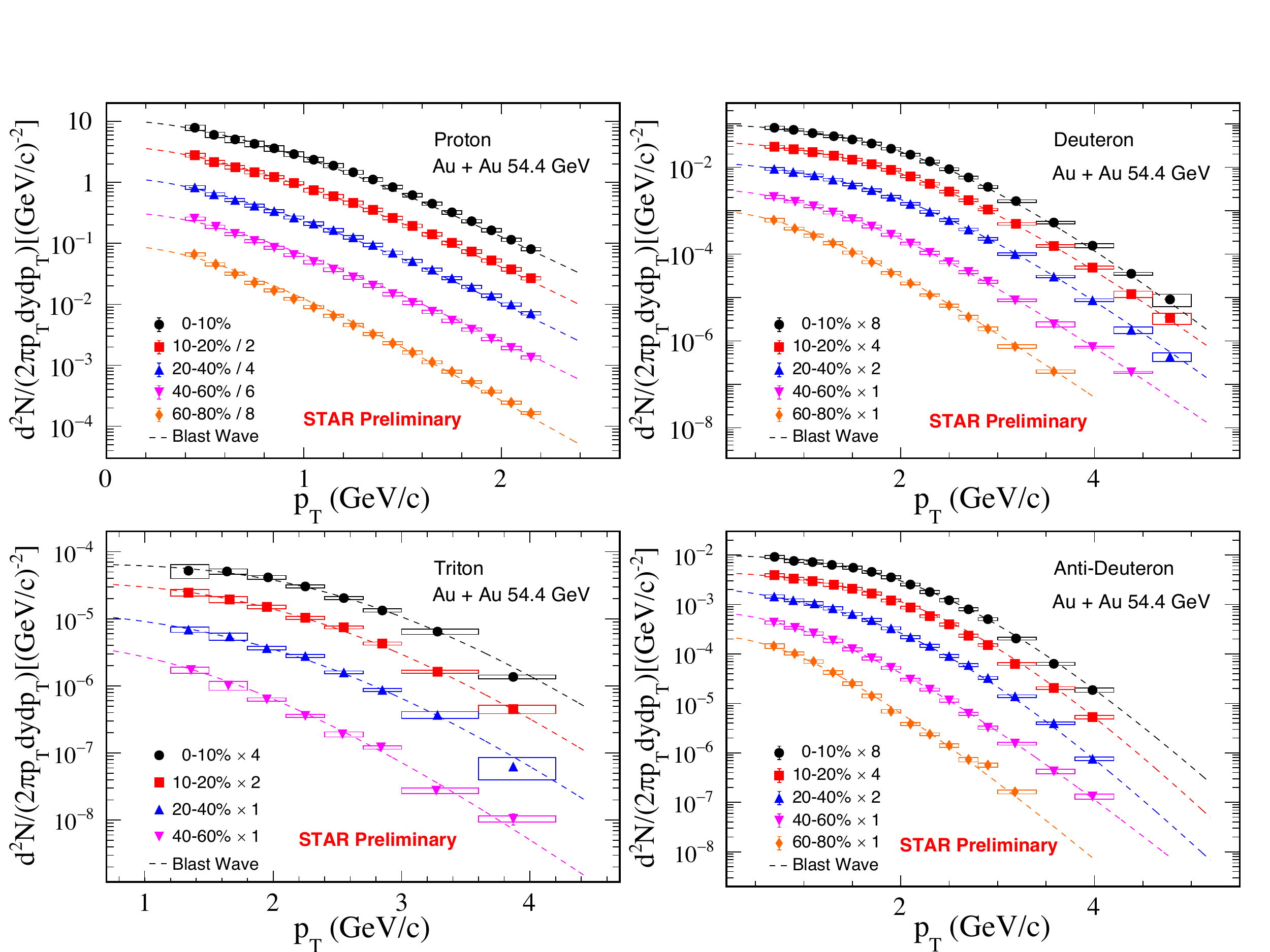}
		\caption{\label{spectra}Transverse momentum spectra of $t$ (left) from BES-I and high-statistical data set at 54.4 GeV (right) measured at midrapidity in Au+Au collisions for different collision centralities. The dashed lines are the individual fits to the data with blast-wave functions. The vertical lines and boxes show the statistical and systematic errors, respectively.}
	\end{figure}
	Fig.~\ref{spectra} left panel shows midrapidity ($|y| <$ 0.5) transverse momentum spectra for tritons in Au+Au collisions at $\sqrt{s_\mathrm{NN}}=$7.7, 11.5, 14.5, 19.6, 27, 39, 62.4, and 200 GeV for 0-10\%, 10-20\%, 20-40\%, and 40-80\% centralities. Right panel shows midraidity transverse momentum spectra for protons ($|y| < $ 0.1), (anti)deuterons ($|y| < $ 0.3), and triton ($|y| < $ 0.5) in Au+Au collisions at $\sqrt{s_\mathrm{NN}}=$ 54.4 GeV for 0-10\%, 10-20\%, 20-40\%, 40-60\%, and 60-80\% centralities~\cite{hui}. The dashed lines are the results of the Blast-Wave fits to each distribution.
	
	\subsection{Coalescence parameters}
\begin{figure}[!htp]
		\centering
		\includegraphics[width=0.7\textwidth]{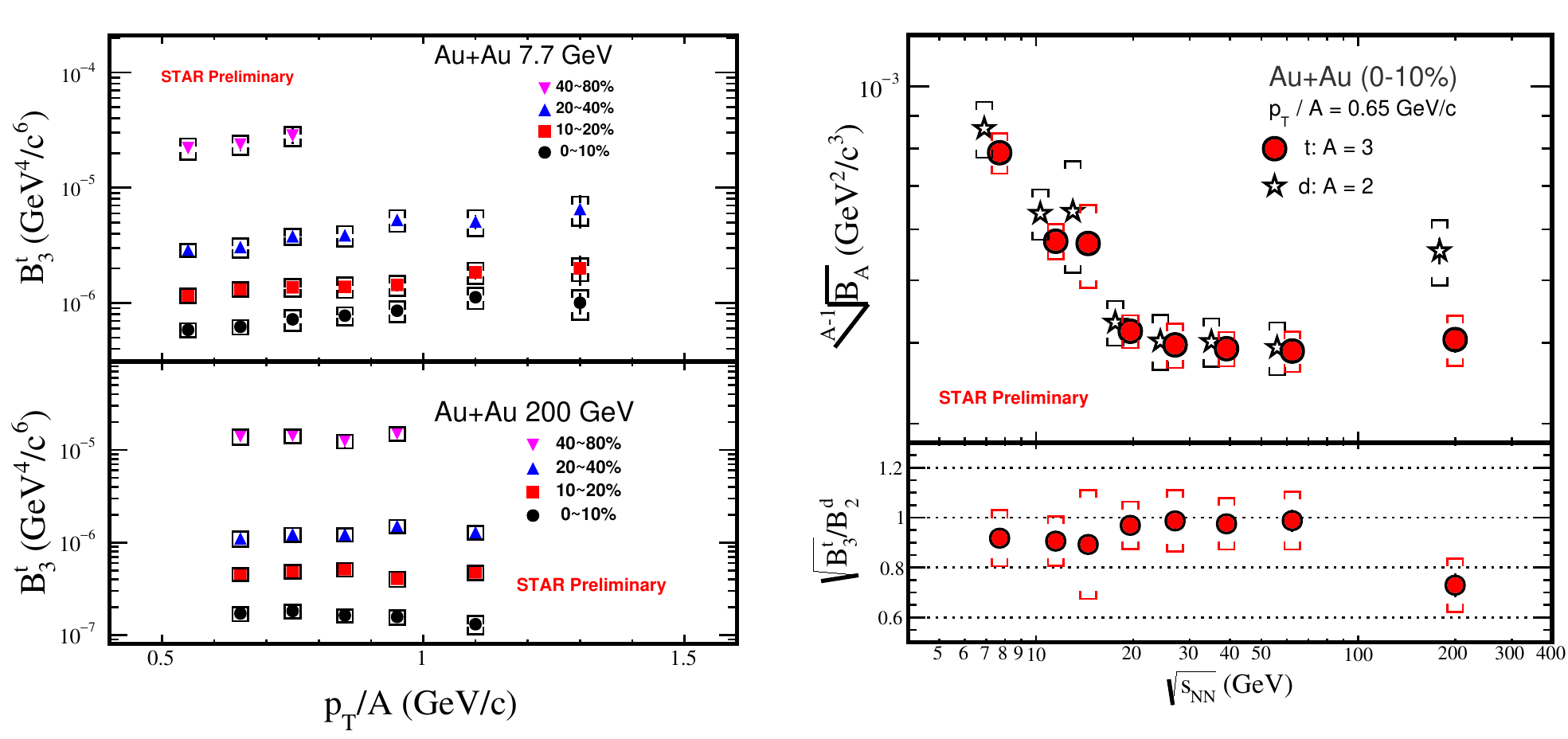}
		\caption{\label{coal}(Left panel) Coalescence parameter $B_3$ as a function of $p_T/A$ for triton measured from 7.7 and 200 GeV collisions in 0-10\%, 10-20\%, 20-40\% and 40-80\% central Au+Au collisions. (Right panel) Coalescence parameter $B_2$ (open star) and $\sqrt{B_3}$ (red dot) as a function of collision energy in 0-10\% central Au+Au collisions. The vertical lines and square brackets show statistical and systematic errors, respectively.}
	\end{figure}

	In the left panel of Fig.~\ref{coal}, the $p_T/A$ dependence of $B_3$ is shown at $\sqrt{s_\mathrm{NN}}=$ 7.7 and 200 GeV for different collision centralities. It is found that the value of $B_3$ increases from central to peripheral collisions and increases with increasing $p_T/A$, which can be explained by the decreasing of source volume.
	In the right panel of Fig.~\ref{coal} we compare the results of $B_2$ and $\sqrt{B_3}$ in the 0-10\% collision centrality at $p_T/A$ = 0.65 GeV/c (This value corresponds to $p_T$=1.2-1.4 GeV/c for $d$, $p_T$=1.8-2.1 GeV/c for $t$). At energies below $\sqrt{s_\mathrm{NN}}$ = 20 GeV, the coalescence parameters $B_2$ and $\sqrt{B_3}$ decrease with increasing collision energy, which implied that the size of the particle-emitting source increases. When $\sqrt{s_\mathrm{NN}}$ $>$ 20 GeV, the decreasing trend seems to change, which might imply a change of the equation of states of the medium in those collisions~\cite{Adam:2019wnb}. The $B_2$ and $\sqrt{B_3}$ are consistent within uncertainties except for $\sqrt{s_\mathrm{NN}}$ = 200 GeV, which might be that the production mechanisms for $d$ and $t$ are different at STAR top energy.
	\subsection{Integral yields and particle ratios}
	The yields of triton and deuteron are obtained by extrapolating the measured $p_T$ range to the unmeasured $p_T$ regions with various parameterizations. The extrapolation is done by individual blast-wave fit for each particle.
	\begin{figure}[!htp]
		\centering
		\centerline{\includegraphics[width=0.98\textwidth]{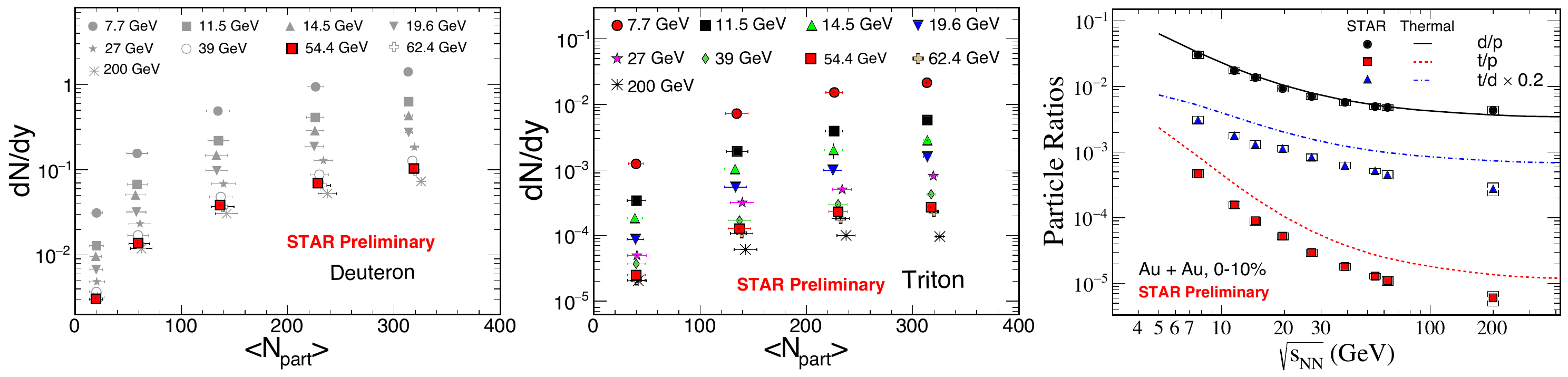}}
		\caption{\label{ratios} Centrality dependence of $dN/dy$ of $d$ (left panel) and $t$ (middle panel) in Au+Au collisions. Energy dependence of $d/p$, $t/p$, and $t/d$ (right panel) ratios for 0-10\% central Au+Au collisions at BES-I energies. The dotted lines are thermal model predictions. The errors represent combined systematic and statistical uncertainties.}
	\end{figure}
	The deuteron integral yields are from the new data sets 54.4 GeV, which confirm the previous measurements at other energies by STAR~\cite{Adam:2019wnb}. We also show the particle ratios of $d/p$, $t/p$, and $t/d$ as a function of collision energy in 0-10\% central Au+Au collisions in the right panel of Fig.~\ref{ratios}. The dashed lines are the thermal model calculations~\cite{Andronic:2010qu}. They describe the ratios of $d/p$ very well but overestimates the $t/p$ and $t/d$ particle ratios.  
	\subsection{Yield ratio of light nuclei}
\begin{figure}[tbh]
		\centering
		\includegraphics[width=0.48\textwidth]{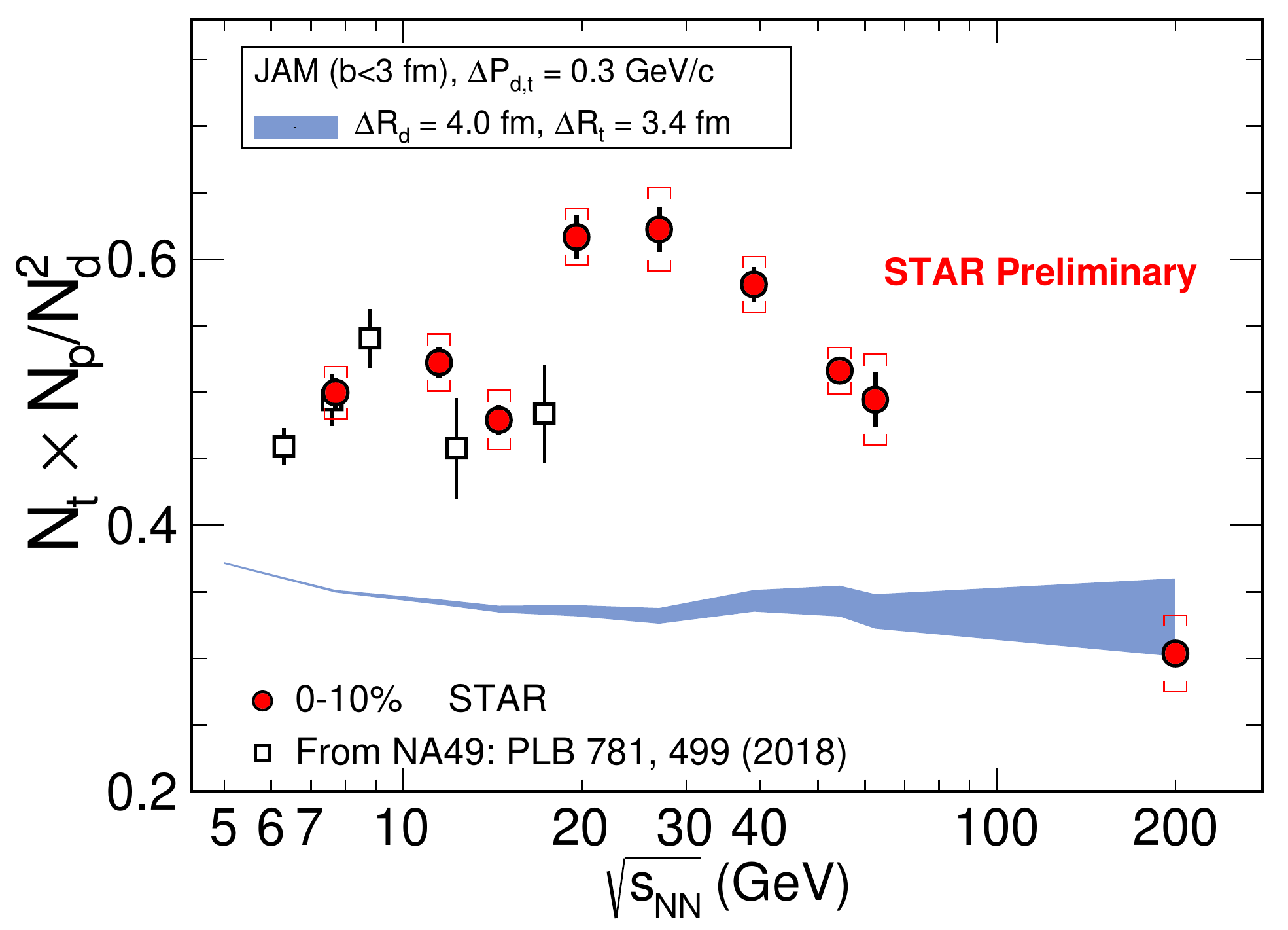}
		\caption{\label{deltan}Energy dependence of the light nuclei yield ratio $N(t)$$\times$$N(p)$/$N^2(d)$ in central ($0-10\%$) heavy-ion collisions. The red solid circles are the results measured in central ($0-10\%$) Au+Au collisions at BES energies by the STAR experiment and the open squares are the results calculated from the Pb+Pb data of NA49 experiment~\cite{Sun:2018jhg}. The blue band represents the results of central Au+Au collisions ($b<3$ fm) from the JAM model calculations~\cite{Liu:2019nii}.}
	\end{figure}

	In Fig.~\ref{deltan}, we show the energy dependence of yield ratio of light nuclei ($N(t)$$\times$$N(p)$/$N^2(d)$), which is directly related to the neutron density fluctuation, in 0-10\% central Au+Au collisions. Non-montonic energy dependence is observed with a peak around 20-30 GeV. The light nuclei yield ratios measured by STAR experiment below 20 GeV are consistent with the results calculated from NA49 experiment~\cite{Sun:2018jhg}. The results from the high statistic Au+Au collisions at 54.4 GeV follow the trend very well. The maximum of this non-monotonic behavior might indicate that the density fluctuations become strongest at this energy region. 
	Since there is no critical physics implemented in the JAM model, the results of central ($b<3$ fm) Au+Au collisions from JAM model is also plotted as blue band in Fig.~\ref{deltan} for comparison. We found that the JAM model results show a flat energy dependence of yield ratio and cannot describe the data~\cite{Liu:2019nii}. To provide a definite physics conclusion on this non-monotonic structure, we still need further studies, especially the precise experimental measurements and theoretical understanding.
	\section{Conclusions}
	We presented STAR results of $d$ and $t$ production in heavy-ion collisions at $\sqrt{s_\mathrm{NN}}=$7.7-200 GeV. The values of $B_{2}$ and $\sqrt{B_3}$ from central collision show the same trends except for 200 GeV and they seem to reach a minimum around 20 GeV, indicating a change in the equation of state. The thermal model cannot describe the triton production, which is still an open question. Non-monotonic energy dependence of the light nuclei yield ratio is observed for 0-10\% central Au+Au collisions with a peak around 20-30 GeV. The ratio below 20 GeV are consistent with the results calculated from NA49 experiment. The results from JAM model, without physics of phase transition and critical point, show flat energy dependence and cannot describe the experimental data. To make definite conclusion, we still need dynamical modelling of heavy-ion collisions with a more realistic equation of state. 
	
	\section*{Acknowledgements}
	This work is supported by the National Key Research and Development Program of China (2018YFE0205201),  the National Natural Science Foundation of China (No.11828501, 11575069, 11890711 and 11861131009).

	\label{}
	
	
	
	
	
	\bibliographystyle{elsarticle-num}

\begin{thebibliography}{00}
		
		\bibitem{Fukushima:2013rx} 
		K.~Fukushima and C.~Sasaki,
		``The phase diagram of nuclear and quark matter at high baryon density,''
		Prog.\ Part.\ Nucl.\ Phys.\  {\bf 72}, 99 (2013)
		\bibitem{Gutbrod:1988gt} 
		H.~H.~Gutbrod, A.~Sandoval et al. 
		``Final State Interactions in the Production of Hydrogen and Helium Isotopes by Relativistic Heavy Ions on Uranium,''
		Phys.\ Rev.\ Lett.\  {\bf 37}, 667 (1976).
		\bibitem{Csernai:1986qf} 
		L.~P.~Csernai and J.~I.~Kapusta,
		``Entropy and Cluster Production in Nuclear Collisions,''
		Phys.\ Rept.\  {\bf 131}, 223 (1986).
		\bibitem{Sun:2018jhg} 
		K.~J.~Sun, L.~W.~Chen, C.~M.~Ko, J.~Pu and Z.~Xu,
		``Light nuclei production as a probe of the QCD phase diagram,''
		Phys.\ Lett.\ B {\bf 781}, 499 (2018)
		\bibitem{hui}
		H. Liu (For the STAR Collaboration). Quark Matter 2019 Poster ID: 389.
		
		\bibitem{Anderson:2003ur} 
		M.~Anderson {\it et al.},
		``The Star time projection chamber: A Unique tool for studying high multiplicity events at RHIC,''
		Nucl.\ Instrum.\ Meth.\ A {\bf 499}, 659 (2003)
		\bibitem{Llope:2005yw} 
		W.~J.~Llope,
		``The large-area time-of-flight upgrade for STAR,''
		Nucl.\ Instrum.\ Meth.\ B {\bf 241}, 306 (2005).
		\bibitem{Adam:2019wnb} 
		J.~Adam {\it et al.} [STAR Collaboration],
		``Beam energy dependence of (anti-)deuteron production in Au + Au collisions at the BNL Relativistic Heavy Ion Collider,''
		Phys.\ Rev.\ C {\bf 99}, no. 6, 064905 (2019)
		\bibitem{Andronic:2010qu} 
		A.~Andronic, P.~Braun-Munzinger, J.~Stachel and H.~Stocker,
		``Production of light nuclei, hypernuclei and their antiparticles in relativistic nuclear collisions,''
		Phys.\ Lett.\ B {\bf 697}, 203 (2011)
		\bibitem{Liu:2019nii} 
		H.~Liu, D.~Zhang, S.~He, N.~Yu and X.~Luo,
		``Light Nuclei Production in Au+Au Collisions at $\sqrt{s_{\mathrm{NN}}}$ = 5-200 GeV from JAM model,''
		arXiv:1909.09304 [nucl-th].
		
		
	\end{thebibliography}
	
	

\end{document}